# A room-temperature electrical-field-enhanced ultrafast switch in organic microcavity polariton condensates

Jianbo De,[1,2] Xuekai Ma,[3] Fan Yin,[2] Jiahuan Ren,[2] Jiannian Yao,[2] Stefan Schumacher,[3,4] Qing Liao,[1,*] Hongbing Fu[1,*] Guillaume Malpuech,[5] Dmitry Solnyshkov,[5,6,*]

[1]Beijing Key Laboratory for Optical Materials and Photonic Devices, Department of Chemistry, Capital Normal University, Beijing 100048, People's Republic of China

[2]Institute of Molecule Plus, Tianjin University, and Collaborative Innovation Center of Chemical Science and Engineering (Tianjin), Tianjin 300072, P. R. China

[3]Department of Physics and Center for Optoelectronics and Photonics Paderborn (CeOPP), Universität Paderborn, Warburger Strasse 100, 33098 Paderborn, Germany

[4]Wyant College of Optical Sciences, University of Arizona, Tucson, Arizona 85721, United States

[5]Institut Pascal, PHOTON-N2, Université Clermont Auvergne, CNRS, Clermont INP, F-63000 Clermont-Ferrand, France

[6]Institut Universitaire de France (IUF), 75231 Paris, France

**ABSTRACT:** Integrated electro-optical switches are essential as one of the fundamental elements in the development of modern optoelectronics. As an architecture for photonic systems exciton polaritons, that are hybrid bosonic quasiparticles that possess unique properties derived from both excitons and photons have shown much promise. For this system, we demonstrate a significant improvement of emitted intensity and condensation threshold by applying an electric field to a microcavity filled with an organic microbelt. Our theoretical investigations indicate that the electric field makes the excitons dipolar and induces an enhancement of the exciton-polariton interaction and of the polariton lifetime. Based on these electric field induced changes, a sub-nanosecond electrical-field-enhanced polariton condensate switch is realized at room temperature, providing the basis for developing an on-chip integrated photonic device in the strong light-matter coupling regime.

**Introduction**

Functional photonic elements are the optical analogue of electronic devices that are central to the development of modern optoelectronics.[1-3] In particular, integrated switches provide the fundamental building blocks for on-chip ultrafast optical connection networks, as well as signal processing.[4, 5] A number of approaches to steer the photonic degree of freedom based on thermal, mechanical, and optical control have been reported.[6] In this context, controlling the optical signal with a static electric field is of a particular interest, because of the ensuing high tunability with external stimuli, compatibility with electronics and the potential for GHz-level switching speeds.[4, 7-11] However, with photons being intrinsically chargeless and massless particles, the direct action of an electric field associated primarily with the effective interaction of photons is weak, which is unfavorable for practical applications. Exciton polaritons (EPs) have emerged as an important platform for exploring new many-body physics and optoelectronic devices that bridge electronic and photonic systems,[12-14] such as polariton gates,[15] transistors,[16] ultrafast switches,[17-18] and memory elements.[19] EPs are half-light half-matter quasiparticles resulting from the strong coupling between excitons and photons in a microscale cavity. The photonic component enables their propagation with high speed and controllable direction over long distances.[20, 21] The excitonic component brings in strong polariton-polariton interactions, giving rise to large nonlinearities.[22-25]

Organic materials are particularly well suited for studies of the strong coupling regime because of their excitons' large oscillator strength. They are ideal to create EPs and achieve polariton condensation at room temperature due to the large binding energy of Frenkel excitons.[26-29] In organic semiconductor microcavities, the molecular vibrations, the equivalent of optical phonons in inorganic ones, are dominantly responsible for the relaxation of polaritons.[16, 30, 31] However, due to the weak Coulomb interaction of the tightly bound Frenkel excitons in organic microcavities, the polariton-polariton and exciton-polariton interactions are usually significantly weaker in comparison with those in inorganic microcavities, which leads to inefficient polariton relaxation to the ground state and weak nonlinearity in the coherent condensate.[29] Therefore, the

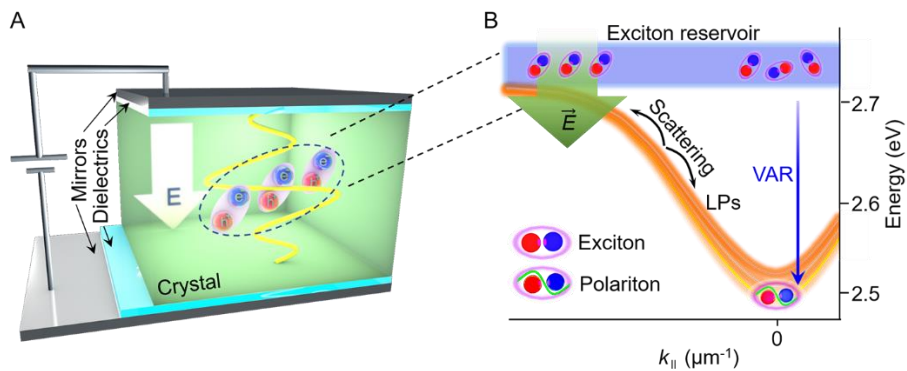

**Figure 1. Schematic diagrams of the electrically contacted microcavity and the dispersions of the lower polariton branches.** (a) The microcavity consists of a DPAVBi microbelt between 200-nm SiO$_2$ spacers that is sandwiched between two silver reflectors. Strong coupling of the cavity modes (indicated by the yellow wave) and the bottom of the excitonic band (reservoir) at 2.75 eV (indicated by blue area) results in multiple lower polariton branches (orange dispersions in (b)). (b) Photogenerated excitons excited by a pulsed laser of 3.1 eV quickly relax to the exciton reservoir (purple area). The direct relaxation to the polaritonic ground state from the exciton reservoir through the VAR channel is indicated by the blue arrow. The applied electric field enhances the polariton-polariton (mediated by their excitonic parts) interaction which leads to efficient stimulated scattering and rapid population of the polaritonic ground state through the polariton-polariton scattering mechanism.

enhancement of the polariton-polariton interaction strength in organic microcavities is much sought after and would lead to strong nonlinearities in polariton systems that are required for the realization of efficient functional devices.[32-34] Application of an electric field provides an alternative approach to manipulate polariton nonlinearities. The electric field aligns electrons and holes making the exciton dipolar and may enhance the stimulated in-scattering of polaritons into the condensate, leading to a reduction of the polariton condensation threshold.[11, 34]

Here, we adopt the electric field strategy and demonstrate significant enhancement of the polariton interaction in an organic microbelt-filled microcavity. It is known that under non-resonant optical excitation without the participation of the electric field, polaritons in organic microcavity condense through vibration-assisted relaxation (VAR) mechanism. The presence of the electric field enhances the polariton-polariton scattering, mediated by their excitonic parts, which then also contributes significantly to the condensation process. As a consequence, the condensation threshold can be reduced and the emission intensity of the polariton condensate is significantly enhanced, with the largest gain of ∼ 37.9 dB observed in the present work, accompanied by a significant blueshift. The enhanced nonlinearity supports the polaritons to condense in different lower polariton branches. Based on that, we realize the operation of a sub-nanosecond electrical-field-enhanced polariton switch at room temperature. The design methodology presented here can be readily generalized to other ultrafast polariton-based functional optoelectronic components and serves as an attractive candidate for on-chip integrated switches for photonic circuits in the strong light-matter coupling regime.

**Result and discussion**

The electrically contacted microcavity used in this work is illustrated schematically in Fig. 1a. The 8.6-μm 4,4'-bis[4-(di-p-tolylamino)styryl]biphenyl (DPAVBi) microbelt (Fig. S1), prepared by a facile reprecipitation method, is sandwiched between two silver reflectors with the thickness of 150 nm and 35 nm,[29, 35] respectively (a detailed description is given in the Supplementary material). The parallel silver reflectors also double as the electrodes to apply the electric field uniformly. In order to prevent electron conduction caused by direct contact of the DPAVBi microbelt with the metallic silver electrodes, 200-nm thick SiO$_2$ films are deposited as transparent insulating spacers between them.

When this microcavity is excited by a non-resonant pulsed laser, the photogenerated excitons quickly relax to the bottom of the excitonic band and form an exciton reservoir (purple area in Fig. 1b at 2.754 eV). These excitons can strongly couple to the cavity photon modes whose polarization is parallel to the belt length direction (Fig. S2), forming polarization-dependent EPs (Fig. S3). Subsequently, the polariton ground state is directly populated by the excitons from the reservoir through the VAR channel (blue arrow in Fig. 1b).[16, 25, 29] To identify the threshold of the polariton condensation in the DPAVBi microbelt-filled microcavity, the excitation-density dependence of polariton emission spectrum was measured. The emission transition from sublinear to superlinear regime at an incident excitation density of ∼41.7 μJ·cm$^{-2}$ ($P_{th0}$) is observed (Fig. S4). Above the threshold, we observe a collapse of the full width at half maximum (FWHM) and an energy blueshift.[29]

To exploit the influence of an electric field on polariton condensation, an additional electric field of 35 kV/cm is applied to this organic microcavity. We measured the excitation-density ($P$) dependence of polariton emission and the typical angle-resolved photoluminescence (ARPL) spectra as shown in Fig. 2. At low pump density of $P$ = 28.2 μJ·cm$^{-2}$, the uniform photoluminescence (PL) emission distributes are the entire detected angular range, following the simulated polariton dispersions (Fig. S3-e), such as LP$_{15}$, LP$_{16}$ and LP$_{17}$ evidencing that the ARPL signals originate from the polariton emission (Fig. 2a). Along with increasing the pump density to $P$ = 42.6 μJ·cm$^{-2}$, the polaritons condense at the bottom of LP$_{17}$ within the angle of around $\theta$ = ±2°, whereas the LP$_{16}$ and LP$_{18}$ polariton emissions remain relatively weak and below the threshold (Fig.

2b). Figure 2c shows the integrated PL intensity (black squares)

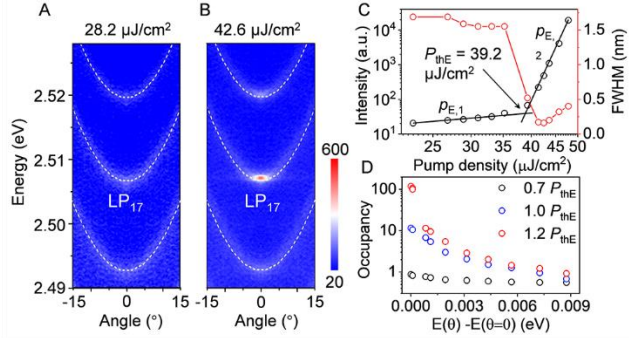

**Figure 2. Polariton condensation with the electric field of 35 kV/cm.** (a) (b) ARPL spectra at different pump densities. (c) The integrate PL intensity (black squares) and the line width (red squares) of the $LP_{17}$ emission as a function of the pump density. (d) Polariton occupancy in the ground state and excited state at $0.7P_{thE}$, $1.0P_{thE}$, and $1.2P_{thE}$.

and FWHM (red squares) of the $LP_{17}$ polariton emission as a function of the pump density. A transition from linear to nonlinear regime occurs at the excitation fluence of 39.2 µJ·cm$^{-2}$, which is the condensation threshold in the presence of the electric field ($P_{thE}$). The intensity dependence is separately fitted to a power law $x^p$, with $p_{E,1}$ = 1.13 ± 0.05 and $p_{E,2}$ = 30.8 ± 0.6 below and above the threshold. This condensation threshold is clearly reduced by about 6.0% in comparison with that without applying the electric field. Meanwhile, the FWHM of the $LP_{17,\theta=0°}$ PL is drastically narrowed from 1.73 nm (below the threshold) to 0.2 nm (above the threshold). Figure 2d shows the occupation of $LP_{17}$ below and above the threshold. As the pump density increases, the polariton occupation of the system transits from Boltzmann to Bose-Einstein distribution with a large number of particles in the ground state, which is a typical feature of polariton condensation.[22]

To further study the influence of the electric field on the polariton condensation, we investigate systematically the polariton emission for different applied electric fields. We initially fix the pump density at 40.1 µJ·cm$^{-2}$ (0.96 $P_{th0}$) for the optical excitation. In this case, no condensation is observed without the electric field, as shown in Fig. 3a. In sharp contrast, the clear polariton condensation at $LP_{17}$ is observed at the electric field of 35 kV/cm (Fig. 3b). We measure the PL spectra and calculate their gain [g = 10×lg($I_E/I_0$)] at different electric field strengths (black dots in Fig. 3e). With the increase of the electric field, the PL intensity is significantly enhanced, similar to the values achieved at higher excitation powers for zero field, and reaches the maximum of 4.87 dB near 35 kV/cm. We find the transition from spontaneous emission to polariton condensation at an electric field of about 19 kV/cm. Simultaneously, the obvious blueshift of the condensate energy is also observed when the electric field changes from 19 kV/cm to 60 kV/cm (black dots in Fig.3f and 3g). Therefore, the presence of the electric field induces stronger nonlinearities and polariton-polariton interaction and results in the enhancement of PL intensity and the reduction of the condensation threshold.

More surprisingly, although the PL intensity of $LP_{16}$ exhibits a gradual decrease as the electric field is increased above 35 kV/cm, the polariton condensation starts to build up at the $LP_{16}$ branch (Fig. 3c) and its intensity rises sharply (blue dots in Fig. 3e). When the electric field exceeds 80 kV/cm, the polariton condensation jumps to the $LP_{15}$ branch along with a significant enhancement of the PL intensity with gain as high as 37.9 dB (red dots in Fig. 3e), as depicted by the cyan lines in Fig. 3b-d. The obvious spectral blueshift can also be seen in two branches of $LP_{16}$ and $LP_{15}$ (Fig. 3f), which suggests that there are strong exciton-polariton and polariton-polariton interactions in this electrically-driven system, inducing strong nonlinearities of organic polaritons. It is note that some branches are not resonant with vibrons applying the electric field (Fig. S5). We also measure the condensation threshold as a function of the electric field. As shown in Fig. 3g and S6a, the reduction of the condensation threshold up to 30.9% (from 41.7 ± 0.7 to 28.8 ± 1.25 µJ·cm$^{-2}$) is achieved at the maximum applied electric field of 118 kV/cm. Generally, a single-step exciton-vibron scattering is predominantly responsible for the population of the polariton state from the exciton reservoir.[16, 31] In our experiments, the energy of $LP_{15}$ is far away from the spontaneous 0-1 PL transition at 2.51 eV. Therefore, the PL enhancement of the $LP_{15}$ branch is dominantly contributed from the polariton-polariton scattering, because the electric field can intensify the polariton-polariton interactions and consequently promote the scattering efficiency to the polariton ground state. Indeed, we carefully investigate the intensity dependence of power law fitted by $x^p$ under the different electric fields and the obtained $p_1$ below threshold are shown in Fig. S6b. The parameter $p_1$ exhibits significant increase with the electric field: $p_1$ increases from 0.69 ± 0.04 without electric field to 1.57 ± 0.06 under electric field of 118 kV/cm. This superlinear dependence strongly supports that the exciton-exciton interactions are enhanced by applying electric field and play an important role in the relaxation.

It has been discussed previously that the exciton-exciton interaction $g_{XX}$ in inorganic semiconductor microcavities can be enhanced by an electric field because it controls the interactions between both indirect excitons and direct excitons[36, 37] and influences the radiative lifetime and dipole moment of excitons.[38] In organic semiconductor microcavities, the main reason for the enhancement of the exciton-exciton interaction with the increase of the electric field is that electrons and holes can be shifted to opposite directions in an electric field which provide them a dipolar nature enhancing their interaction $g_{XX}$. Since the polariton-polariton scattering rate is proportional to $|\beta|^2 g_{XX}$,[39, 40] where $|\beta|^2$ denotes the exciton fraction (Hopfield coefficient) of a polariton and it reduces as the detuning of the cavity photons and excitons decreases (for example, in our sample, $|\beta|^2$ = 0.948 for $LP_1$ and $|\beta|^2$ = 0.313 for $LP_{17}$), when the condensates jump to higher LP branches as the electric field becomes stronger, the polariton-polariton interaction is strengthened further as can be seen in Fig. 3e-g.

To theoretically understand the influence of the electric field on the density of the polariton condensate, we evalu-

ate a set of simplified rate equations based on semi-classical Boltzmann kinetic equations, neglecting the

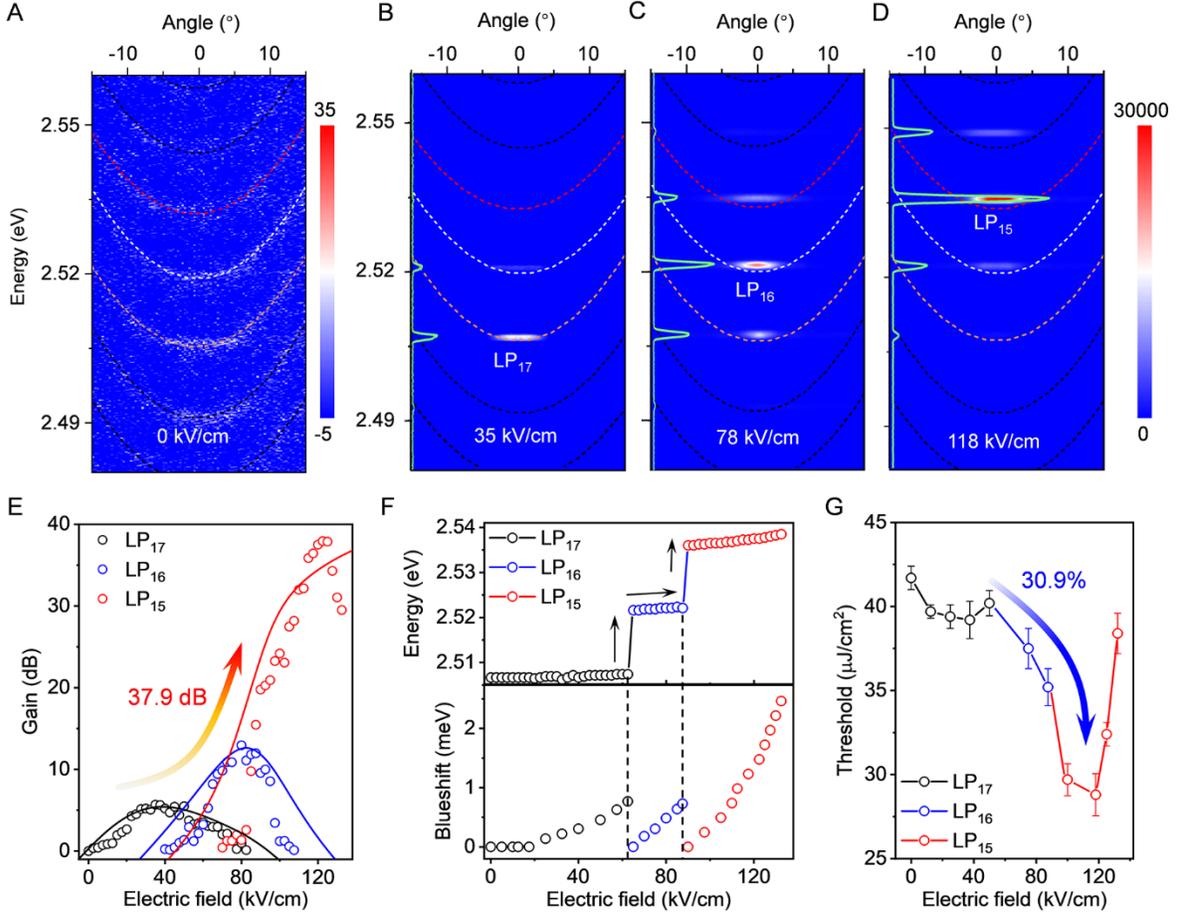

**Figure 3. Polariton condensation at different LP branches**. ARPL spectra measured at 0 (a), 35 (b), 78 (c), and 118 kV/cm (d). Dependence of emitted-intensity gain (dots are experimental results and lines are theoretical results) (e), energy, blueshift (f), and condensation threshold (g) of polariton emission on the strength of the electric field.

momentum-dependence of the various scattering processes:[38, 41-43]

$$\frac{dn_i}{dt}=W_i n_r^2(1+n_i)x_i-(\Gamma_p(1-x_i)+\Gamma_p x_i)n_i \quad (1)$$

$$\frac{dn_r}{dt}=P-\Gamma_{Xr}n_r-\sum_i W_i n_r^2(1+n_i)x_i \quad (2)$$

Here, $n_r$ is the reservoir population and $n_i$ are the populations of the 3 polariton branches; $\Gamma_x$ and $\Gamma_p$ are the exciton and the photon decay rates, respectively; $x_i$ are the excitonic fractions of the branches found from the strong coupling Hamiltonian; $W_i$ are the scattering rates from the reservoir to the polariton branches, based on the exciton-exciton scattering, that we assume to be dominating at the threshold thanks to the applied electric field. To simulate the pulsed pumping, we take $P=P_0\delta(t)$, which provides an initial reservoir population $n_{r0}$.

The electric field affects the different system parameters differently. The Rabi splitting $V$ and the radiative lifetime of the excitonic reservoir $\Gamma_{xr}$ are proportional to the overlap of the electron and the hole wavefunctions within the exciton, which decays exponentially when the exciton is polarized by an external field, as shown previously in in *ref.* 43. We therefore take $V\sim e^{-E/E}$ and $\Gamma_{xr}\sim e^{-E/E_{sat}}$, where the saturation value of the electric field $E_{sat}$ is a fitting parameter. On the other hand, the increase of the exciton size improves the exciton-exciton scattering rate linearly with the dipole moment induced by the electric field.[44, 45] We therefore take $W_i\sim W_0+\alpha(E/E_{sat})^2$. The consequences of these dependencies are twofold: 1) the increase of the scattering rates increases the efficiency of the relaxation with the applied field for all branches, allowing to pass the condensation threshold for fixed pumping; 2) the decrease of the Rabi splitting reduces the exciton fraction of the lowest branches, making them less favorable for relaxation and leading to the switching to higher branches for higher fields.

The theoretical results regarding the increase of the condensate density caused by the electric field are presented in Fig. 3e (solid lines), which describe well the growth of the emitted-intensity gain at each LP branch observed in experiments and the switching between the branches (from lower to upper). The values of the parameters in Eqs. (1)-(2) used for the calculation of the curves in Fig. 3e can be found in Table S2 in the Supplementary materials. The good agreement between the experiment and the theory supports our interpretation of the experimental results.

In organic systems, the blue shift arises mostly from the phase space filling effects and is therefore directly related to the decrease of the Rabi splitting. The relatively weak blue shifts that we observe (Fig. 3f) indicate a weak modification of the Rabi splitting (less than 10%) and therefore of the excitonic fractions. Since a small modification of the excitonic fractions cannot lead to the strong increase of the laser emission that we observe, we therefore conclude that this enhancement is due to the increase of the interactions. Indeed, our numerical simulations indicate a 10-fold improvement of the interactions, which is relatively modest for excitons with a dipole moment.[45]

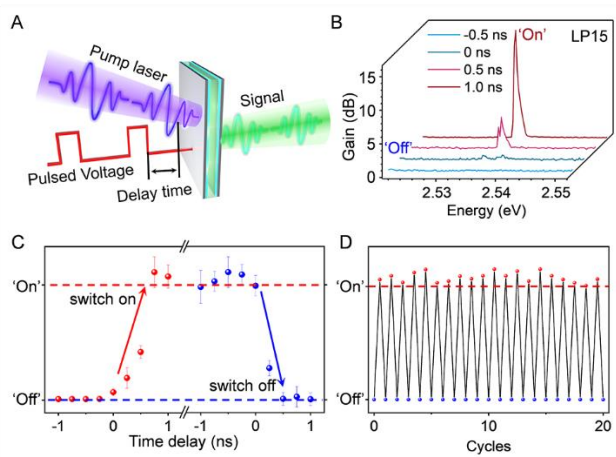

**Figure 4. Characteristics of the polariton switch.** (a) Schematic diagram of a polariton switch. (b) PL spectra of the $LP_{15}$ branch for different delay time. 'On' state is defined as the gain of the $LP_{15}$ branch over 10 dB. (c) 'On/Off' state of the switch device versus the time delay between the pulsed voltage and pump laser. (d) Cyclic 'On/Off' switching behavior, which is changed by continue sweeping time delay for 20 cycles.

Taking advantage of the prominent nonlinearity in the bosonic relaxation, observed for the first time in organic crystal-filled microcavities, we demonstrate that the switching of the polariton condensate can work as a room-temperature ultrafast switch thanks to the inherent ultrafast exciton relaxation dynamics and the short lifetime on a picosecond scale of the polariton condensates. Figure 4a depicts schematically the principle of the switch operation, wherein the 'On' state and 'Off' state of the switch is controlled by two fields, i.e., the pulsed voltage and pulsed laser. Meanwhile, the pulsed field configuration also allows a time delay, which can characterize the temporal behavior of the transition process. Here, we employ a digital delay pulse generator and a voltage amplifier (Scheme S3) to generate a square wave pulsed voltage with an adjustable delay (Fig. S7). Figure 4b shows the PL spectra of $LP_{15}$ at $\theta$ = 0° at different delay time with the pump density fixed at 1.4 $P_{th0}$. If the microcavity is pumped only by the pulsed laser, the gain is 0 dB, which is defined as the 'Off' state. Once the electric field overlaps temporally with the pulsed laser, the gain increases. To accurately characterize the switching time, we defined the 'On' state as the gain of $LP_{15}$ branch exceeds 10 dB. Fig. 4c presents the switching time from 'Off' state to 'On' state and back to off state by scanning the time delay between the pump laser and pulsed voltage. The response time of the switching between on state and off state is determined to be less than 0.75 ns. In fact, we consider that our polariton switch should have a faster response rate because its response time is directly related to the dynamics of excitons and polaritons in the organic microcavity. However, limited by the pulsed voltage resolution, only the sub-nanosecond switch has been realized. Figure 4d shows the robustness of the polariton switch. The switching behavior remains almost the same for all 20 cycles, demonstrating stable switching behavior with low device fatigue.

**Conclusions**

In summary, we demonstrate the enhancement of the polariton-polariton interaction by application of an electric field to an organic crystalline microcavity. This leads to a reduction of the polariton condensation threshold and a dispersion-branch dependent blueshift. The stronger polariton-polariton interaction provides a new scattering channel besides VAR. Based on these observations, we realize an electric field-controlled ultrafast polariton switch at room temperature. Our results on the one hand provide further insight into organic polariton condensation. On the other hand, they promote the advancement of polariton-based functional optical elements for miniaturized optoelectronic circuits and on-chip integration of photonics and electronics.

AUTHOR INFORMATION

Corresponding Author

**Qing Liao** – *Beijing Key Laboratory for Optical Materials and Photonic Devices, Department of Chemistry, Capital Normal University, Beijing 100048, People's Republic of China;*
Email: liaoqing@cnu.edu.cn
**Hongbing Fu** – *Beijing Key Laboratory for Optical Materials and Photonic Devices, Department of Chemistry, Capital Normal University, Beijing 100048, People's Republic of China;*
Email: hbfu@cnu.edu.cn
**Dmitry Solnyshkov** – *Institut Pascal, PHOTON-N2, Université Clermont Auvergne, CNRS, Clermont INP, F-63000 Clermont-Ferrand, France; Institut Universitaire de France (IUF), 75231 Paris, France;*
Email: dsolnyshkov@gmail.com

Author Contributions

The manuscript was written through contributions of all authors.

Funding Sources

This work was supported by the National Key R&D Program of China (Grant No. 2018YFA0704805, 2018YFA0704802 and 2017YFA0204503), the National Natural Science Foundation of China (22150005, 22090022, 21833005 and 21873065), the Natural Science Foundation of Beijing, China (KZ202110028043), Beijing Talents Project (2019A23), Capacity Building for Sci-Tech Innovation-Fundamental Scientific Research Funds, Beijing Advanced Innovation Center for Imaging Theory and Technology. The Paderborn group acknowledges support by the Deutsche Forschungsgemeinschaft (DFG) through the collaborative research center TRR142 (project A04, No. 231447078). We also acknowledge the support of the European Union's Horizon 2020 program, through a FET Open research and innovation action under the grant agreement No. 964770 (TopoLight), the ANR Labex Ganex (ANR-11-LABX-0014), and of the ANR program "Investissements d'Avenir" through the IDEX-ISITE initiative 16-IDEX-0001 (CAP 20-25).

## ACKNOWLEDGMENT

The authors thank Dr. H.W. Yin from ideaoptics Inc. for the support on the angle-resolved spectroscopy measurements.

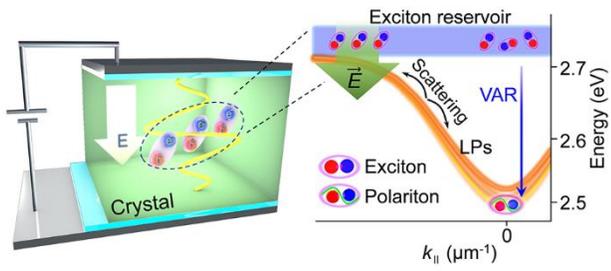

SYNOPSIS TOC (Word Style "SN_Synopsis_TOC").

We demonstrate the enhancement of the polariton-polariton interaction by application of an electric field to an organic crystalline microcavity. This provides a new scattering channel besides VAR further leads to a reduction of the polariton condensation threshold and a dispersion-branch dependent blueshift. Stand on consideration, we realize an electric field-controlled ultrafast polariton switch at room temperature.

# Supplementary material:

# A room-temperature electrical-field-enhanced ultrafast switch in organic microcavity polariton condensates

Jianbo De,[1,2] Xuekai Ma,[3] Fan Yin,[2] Jiahuan Ren,[2] Jiannian Yao,[2] Stefan Schumacher,[3,4] Qing Liao,[1,*] Hongbing Fu[1,*] Guillaume Malpuech,[5] Dmitry Solnyshkov,[5,6,*]

[1]Beijing Key Laboratory for Optical Materials and Photonic Devices, Department of Chemistry, Capital Normal University, Beijing 100048, People's Republic of China

[2]Institute of Molecule Plus, Tianjin University, and Collaborative Innovation Center of Chemical Science and Engineering (Tianjin), Tianjin 300072, P. R. China

[3]Department of Physics and Center for Optoelectronics and Photonics Paderborn (CeOPP), Universität Paderborn, Warburger Strasse 100, 33098 Paderborn, Germany

[4]Wyant College of Optical Sciences, University of Arizona, Tucson, Arizona 85721, United States

[5]Institut Pascal, PHOTON-N2, Université Clermont Auvergne, CNRS, Clermont INP, F-63000 Clermont-Ferrand, France

[6]Institut Universitaire de France (IUF), 75231 Paris, France

**MATERIALS AND METHODS**

**1. The preparation of DPAVBi microbelts**

In our experiment, DPAVBi microbelts were fabricated using facile reprecipitation method, upon injection of 140 μL of a stock solution of DPAVBi in Tetrahydrofuran (THF) into 2 mL of Hexane, then silence at 283 K for 2 hours. Finally, crystals are washed with ethanol and transferred to the substrate for the next characterization.

**2. The preparation of DPAVBi microcavity**

Firstly, we use the metal vacuum deposition system (Amostrom Engineering 03493) to thermally evaporate a silver film with the thickness of 150 ± 5 nm (reflectivity R ≥ 99%) on the glass substrate, the root mean square roughness (Rq) of the silver film in the 5 μm × 5 μm area is 2.23 nm, a 200 ± 5 nm $SiO_2$ layer was deposited using vacuum electron beam evaporate on the silver film with $R_q$ of 2.15 nm, the deposited rates were both 2 Å/s and the base vacuum pressure is $3×10^{-6}$ Torr. This silver/$SiO_2$ film composite structure was placed as a substrate in a horizontal tube furnace for sample deposition. The DPAVBi microbelts were uniformly dispersed on the silver/$SiO_2$ film substrate. Then 200 ± 5 nm $SiO_2$ and 35 ± 2 nm (R ≈ 50%) silver was fabricated to form the microcavity. The 200-nm $SiO_2$ layer is inserted between the silver film and the crystal as an insulating layer to isolate the crystal from the electrode.

**3. Structural and spectroscopic characterization**

As-prepared DPAVBi microbelts were characterized by field emission scanning electron microscopy (FE-SEM, HITACHI S-4800) by dropping on a silicon wafer.

The X-ray diffraction (XRD, Japan Rigaku D/max-2500 rotation anode X-ray diffractometer, graphite monochromatized Cu K$_\alpha$ radiation ($\lambda$ =1.5418 Å)) operated in the 2θ range from 3 to 30°, by using the samples on a cleaned glass slide.

The fluorescence micrograph, diffused reflection absorption and emission spectra were measured on Olympus IX71, HITACHI U-3900H, and HITACHI F-4600 spectrophotometers, respectively. The photoluminescence spectra of isolated single DPAVBi microbelt in microcavity was characterized by using a homemade photoelectric integrated detection system equipped with a 50 × 0.42 NA objective lens (working distance 20.5 mm) (Scheme S1). The second harmonic ($\lambda$ = 400 nm, pulse width 150 femtosecond) of a 1 kHz Ti: sapphire regenerative amplifier was focused to a 50-μm diameter spot to excite the selected single DPAVBi on a two-dimensional (2D) movable table. A sourcemeter (Keithley 2400) equipped with probes is applied to electrical measurement.

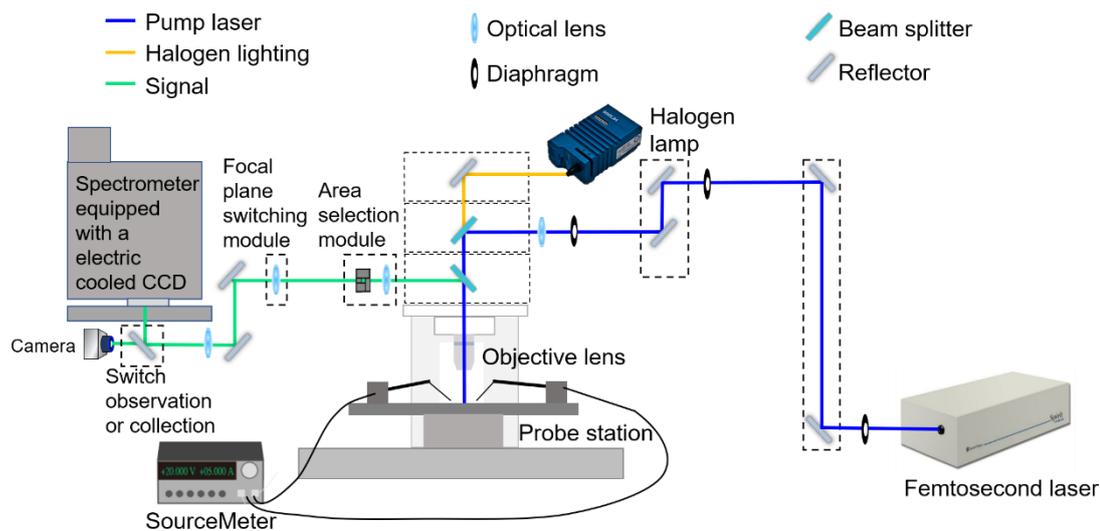

**Scheme S1.** Schematic demonstration of the experimental setup for the photoelectric characterization. In order to maintain stability, the whole set of equipment is placed on the optical platform. The blue line represents the pump laser, the yellow line represents the white light path origin from a halogen lamp and the green line is the collection and observation path.

## 4. Angle-resolved spectroscopy characterization

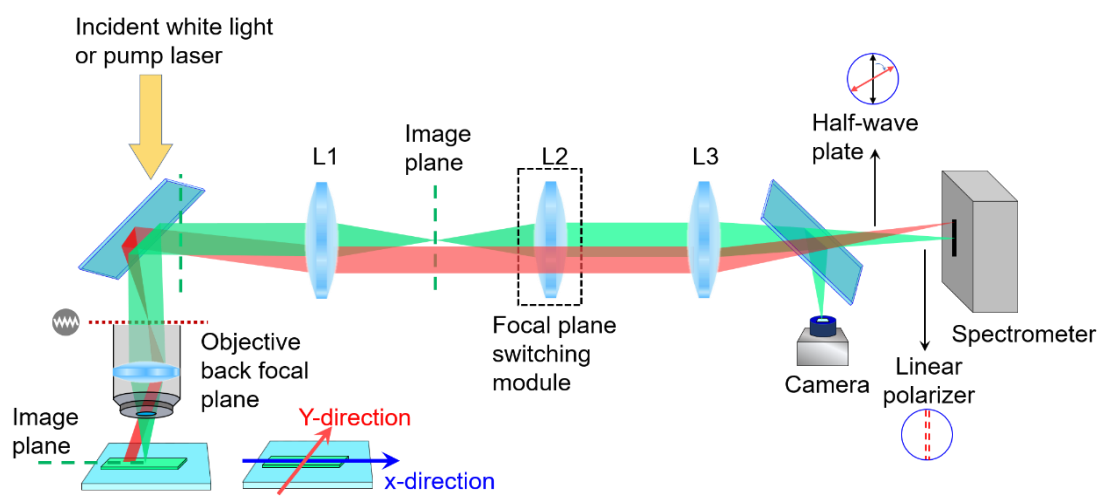

**Scheme S2.** Experimental setup allowing to obtain polarization-resolved spectroscopy. L1-L3: lenses. The red beam traces the optical path of the reflected light from the sample at a given angle. The green beam traces the optical path of the image plane.

The angle-resolved PL spectroscopy was performed at room temperature by the Fourier imaging using a 50× objective lens of a NA 0.42, corresponding to a range of collection angle of ±15° (Scheme S2). The angle-resolved reflectivity was collected by a 100× objective lens of a NA 0.95. An incident white light from a Halogen lamp with the wavelength range of 400-700 nm was focused on the area of the microcavity containing a DPAVBi microbelt. The k-space or angular distribution of the reflected light (red light path in scheme S2) was located at the back focal plane of the objective lens. The L2 is a focal plane switching lens, used for switching the right focal plane of L3 between image plane and back focal plane. When the L2 is out, Lenses L1 and L3 formed a confocal imaging system together with the objective lens, by which the k-space light distribution was imaged at the right focal plane of L3 on the entrance slit of a spectrometer equipped with an electric-cooled CCD. When the L2 is inserted, the right focal plane of L3 is converted to image plane which can be recorded by camera.

In order to investigate the polarization properties, we placed a linear polarizer, a half-wave plate in front of spectrometer to obtain the polarization state in the horizontal-vertical (0° and 90°)

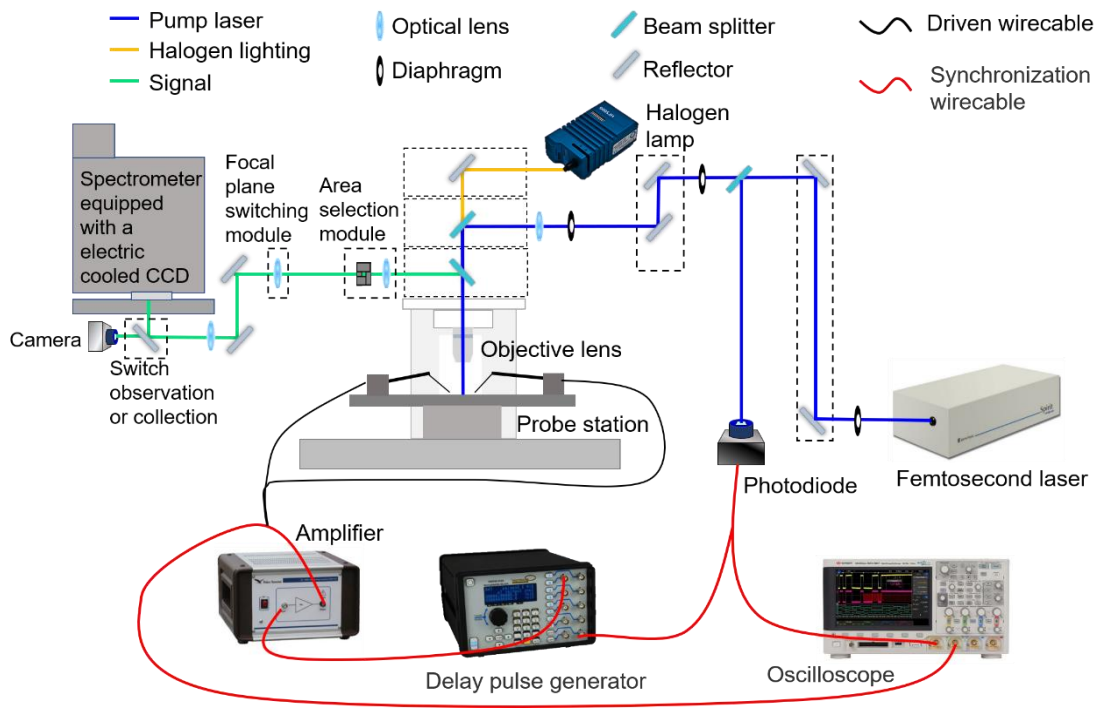

**Scheme S3.** Schematic demonstration of the response time measurement. The blue line represents the pump laser, the yellow line represents the white light path origin from a halogen lamp, the green line is the collection and observation path, the dark line is driven wire-cable and the red line is synchronization wire-cable. In order to realize the synchronization of the electric pulse and the laser pulse, we use a beam splitter to reflect part of the pump light onto the photodiode and convert it into an electric signal, then input this signal as a trigger signal (Sig 1) to the delayed pulse generator and monitored with the oscilloscope (input channel 1). The 1kHz square wave signal generated by the delayed pulse generator (accuracy of 250 ps) is amplified by the amplifier and used to drive the device and monitored by the oscilloscope (input channel 2) at the same time.

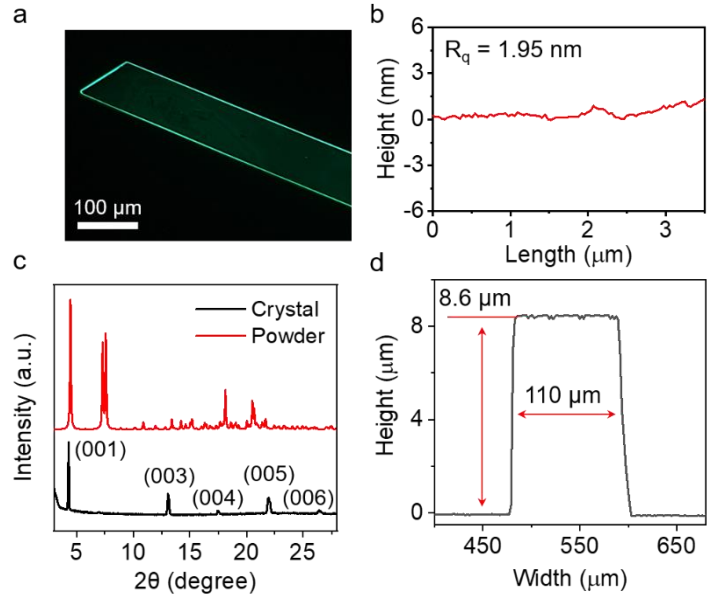

**Figure S1.** (a) Fluorescence image of the as-fabricated organic microcavity. (b) The surfaces roughness of microbelts measured by AFM with Rq = 1.95 nm. (c) X-ray diffraction (XRD) spectrum of DPAVBi microbelts and powder. (d) The dimensions of as-prepared DPAVBi microbelts measured by the profilometer. The width and thickness are around 110 μm and 8.6 μm, respectively.

The XRD spectrum is dominated by a series of diffraction peaks corresponding to crystal plane (001) with d = 20.5 Å, indicating that DPAVBi molecules adopt a lamellar structure with (001) plane parallel to the substrate. On the basis of crystallographic analysis, DPAVBi molecules are highly ordered uniaxial aligned along [010] direction (long axis of microbelts) preferentially, which leads to that the projection of transition dipole-moment ($\mu$) along the microbelt length direction (denoted as parallel direction $\mu_{//}$) is much larger than that along the microbelt width direction (denoted as vertical direction $\mu_{\perp}$). (Ren J., et al. *Nano Lett.* 2020, **20**, 7550-7557.)

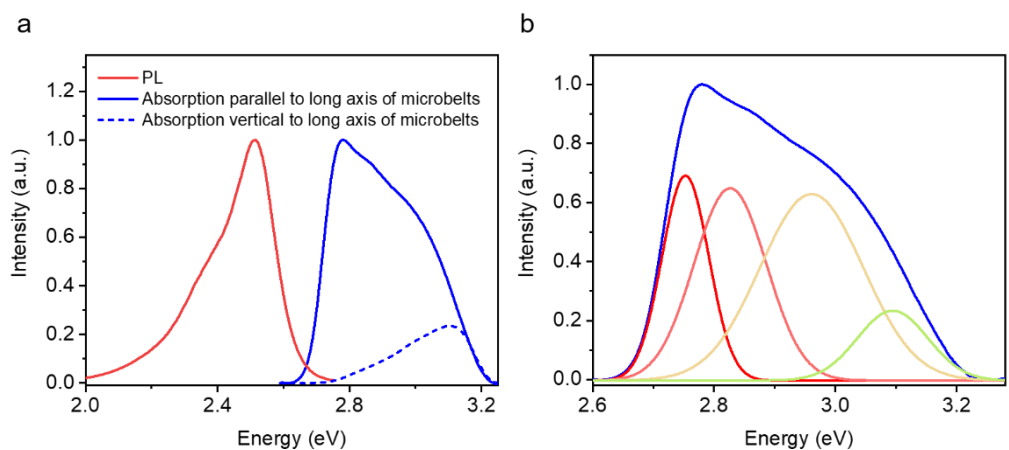

**Figure S2.** Polarization-dependent absorption and PL spectra of a single microbelt. (a) The blue solid line is the absorption spectrum parallel to long axis of microbelts and the blue dash line is vertical to long axis of microbelts. (b) Gaussian split peak fitting to absorption spectra. These distinct polarization-dependent absorption features are consistent with the fact that $\mu_{//} > \mu_{\perp}$ as a result of the highly ordered uniaxial alignment of DPAVBi molecules in single crystalline microbelts. Combined with the distinct H-aggregation characteristic of DPAVBi molecules, polarization-dependent absorption and PL spectrums of single microbelt, an exciton reservoir at 2.754 eV (purple belt in Fig. 1b) was formed by vibrational relaxation, that is, internal conversion of hot excitons (excitons coupled with vibrational energy injected through non-resonant excitation) from high energy excited state to the bottom of the first excited electronic state $S_1$.

## 5. Polariton dispersion

The polariton dispersion in Fig. 1b was calculated by a coupled harmonic oscillator Hamiltonian (CHO) model (S. Kena-Cohen *et al.*, *Phys. Rev. Lett.* 2008, 101, 116401.). The 2×2 matrix in equation (1) below describes the CHO Hamiltonian:

$$\begin{pmatrix} E_{CMn}(\theta) & \Omega/2 \\ \Omega/2 & E_X \end{pmatrix} \begin{pmatrix} \alpha \\ \beta \end{pmatrix} = E \begin{pmatrix} \alpha \\ \beta \end{pmatrix} \quad (1)$$

Where $\theta$ represents the polar angle, $E_{CMn}(\theta)$ is the cavity photon energy of the $n^{th}$ cavity mode as a function of $\theta$, $E_X$ is the exciton 0-0 absorption energy of DPAVBi microbelts at 2.754 eV (450 nm) and $\Omega$ (eV) denotes the coupling. The magnitudes $|\alpha|^2$ and $|\beta|^2$ correspond to the photonic and the excitonic fraction, respectively. We therefore replace the complex broadband absorption of DPAVBi by a single effective resonance which includes the contribution to coupling of all excitonic lines. This approximation should well reproduce the lower polariton mode dispersion because the Rabi splitting values exceed the width of the absorption band.

The dispersion of the $n^{th}$ cavity mode is given by:

$$E_{CMn}(\theta) = E_{cn}\sqrt{1 + \sin^2 \theta} \quad (2)$$

where $E_{cn} = n\frac{\hbar \pi c n_{eff}}{L}$ represents the energy of the $n^{th}$ cavity mode at $\theta = 0°$, with $L$ the cavity thickness. One should note that each photonic mode is characterized by a profile along the growth axis of the cavity and actually couples to a particular exciton distribution of the bulk active media. As a result, one should not consider many modes coupled to a single excitonic resonance, but really that each cavity mode characterized by n (and $\theta$) couples to a single exciton resonance. Diagonalization of this Hamiltonian yields the eigenvalues, $E_{\pm}(\theta)$, which represent the upper and lower polariton (UP and LP) in-plane dispersions (H. Deng *et al.*, *Rev. Mod. Phys.* 2010, **82**, 1489-1537.),

$$E_{\pm}(\theta) = \frac{E_X + E_{CMn}(\theta)}{2} \pm \frac{1}{2}\sqrt{(E_X - E_{CMn}(\theta))^2 + \hbar^2 \Omega^2} \quad (3)$$

We fabricated several microcavities with the crystal thicknesses of 785 nm, 1589

nm, 2322 nm, 4232 nm, and 8800 nm and recorded their reflection spectra in H and V polarization in the range of -60° ~ 60° by using a homemade ARR setup at room temperature (Scheme S1 and S2). The results are displayed on figure S3-(a-e) together with the fit of the dispersion obtained with formula 2-3 The refractive indices in V and H polarization are $n_{eff}$ = 1.592 and $n_{eff}$ = 2.157 respectively. The fit parameters are displayed in table S3-a to S3-e.

Due to the strong absorption signal of the microbelts, it is difficult to detect an effective signal above the reservoir of 2.754 eV in the angle-resolved reflectivity, as this part appears to be grey. Therefore, the upper polariton branches cannot be observed from the angle-resolved reflectivity.

The agreement between experiment and the very basic coupled oscillator model strongly suggests the validity of the polariton picture to describe our experimental data.

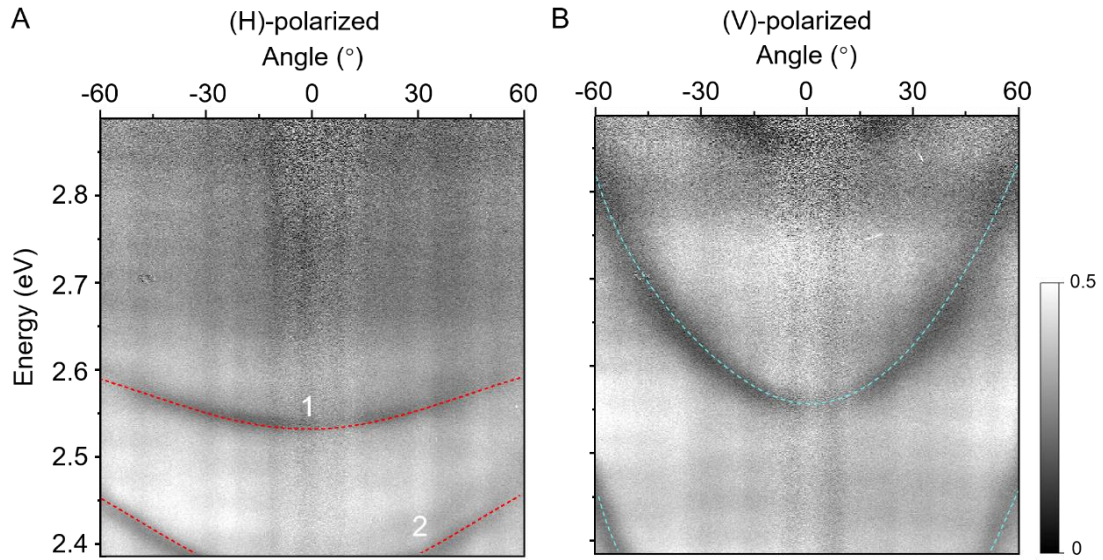

Figure S3-a. The angle-resolved reflectivity of the 785-nm DPAVBi microbelt at (A) V-polarization coupled LP and (B) H-polarization uncoupled cavity mode. The Rabi splitting energy is fitted to be 756 meV.

| coupling mode | $LP_1$ | $LP_2$ |
|---|---|---|
| $E_{\theta=0°}$ (eV) | 2.52498 | 2.35473 |

| detuning (meV) | 230 | -30 |
|---|---|---|
| $|\alpha|^2$ | 0.791861 | 0.46071 |
| $|\beta|^2$ | 0.208139 | 0.53929 |

Table S3-a. The fitted parameters of the 785-nm DPAVBi microbelt.

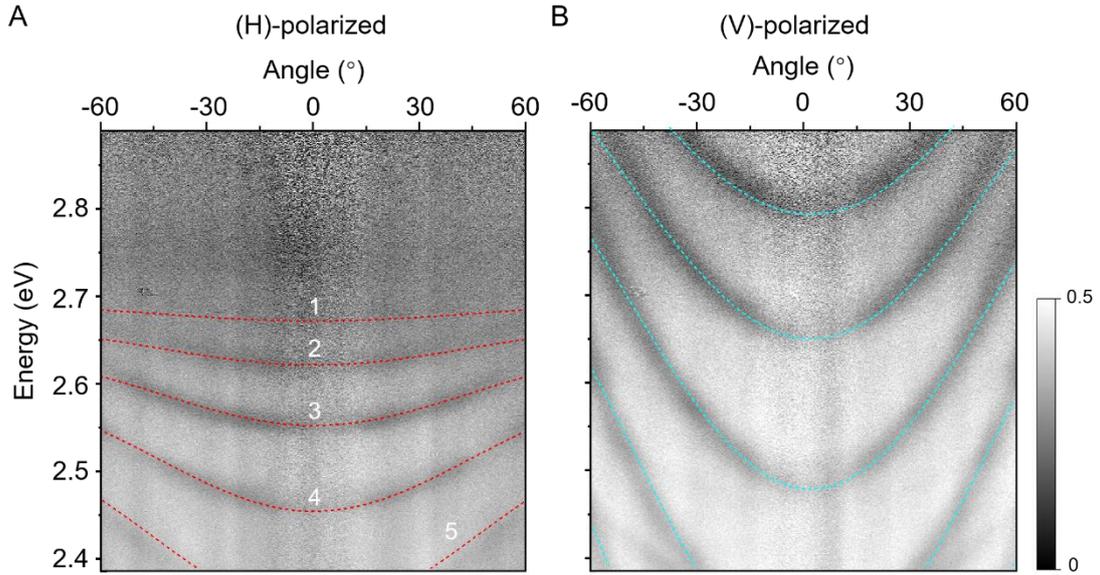

Figure S3-b. The angle-resolved reflectivity of the 1589-nm DPAVBi microbelt at (A) V-polarization coupled LP and (B) H-polarization uncoupled cavity mode. The Rabi splitting energy is fitted to be 610 meV.

| coupling mode | LP$_1$ | LP$_2$ | LP$_3$ | LP$_4$ | LP$_5$ |
|---|---|---|---|---|---|
| $E_{\theta=0°}$ (eV) | 2.67583 | 2.61915 | 2.5454 | 2.44751 | 2.34203 |
| detuning (meV) | 1180 | 480 | 250 | 5 | -180 |
| $|\alpha|^2$ | 0.984092 | 0.942551 | 0.817003 | 0.508274 | 0.24592 |
| $|\beta|^2$ | 0.015908 | 0.057449 | 0.182997 | 0.491726 | 0.75408 |

Table S3-b. The fitted parameters of the 1589-nm DPAVBi microbelt.

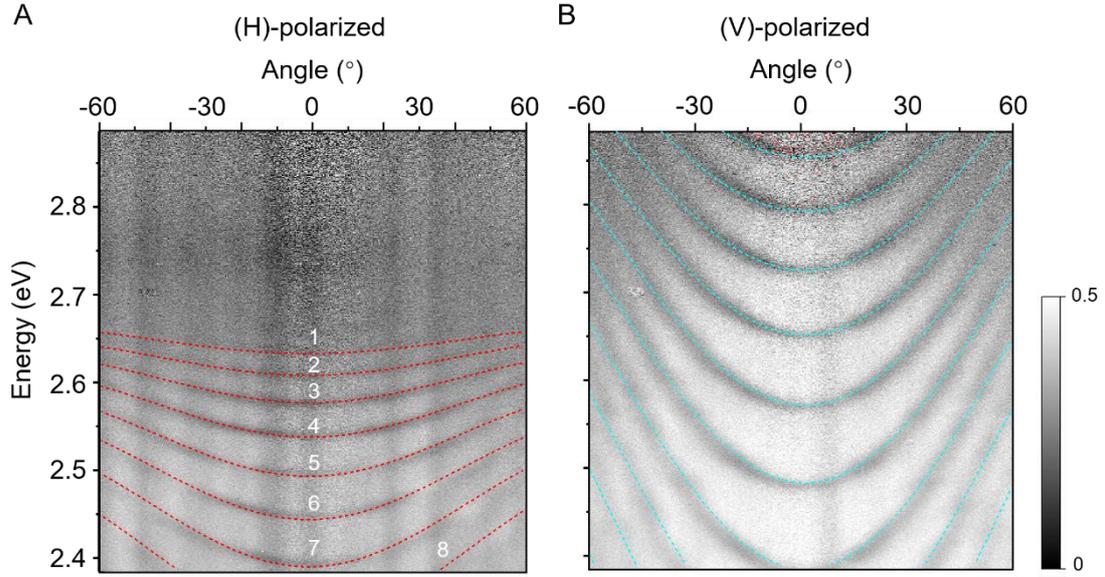

Figure S3-c. The angle-resolved reflectivity of the 2322-nm DPAVBi microbelt at (A) V-polarization coupled LP and (B) H-polarization uncoupled cavity mode. The Rabi splitting energy is fitted to be 602 meV.

| coupling mode | LP$_1$ | LP$_2$ | LP$_3$ | LP$_4$ | LP$_5$ |
|---|---|---|---|---|---|
| $E_{\theta=0°}$ (eV) | 2.63271 | 2.60582 | 2.57263 | 2.53207 | 2.48736 |
| detuning (meV) | 650 | 480 | 330 | 195 | 80 |
| $|\alpha|^2$ | 0.953986 | 0.924012 | 0.869995 | 0.77254 | 0.628898 |
| $|\beta|^2$ | 0.046014 | 0.075988 | 0.130005 | 0.22746 | 0.371102 |
| coupling mode | LP$_6$ | LP$_7$ | LP$_8$ | | |
| $E_{\theta=0°}$ (eV) | 2.43986 | 2.39002 | 2.33314 | | |
| detuning (meV) | -20 | -110 | -201 | | |
| $|\alpha|^2$ | 0.466812 | 0.32793 | 0.22173 | | |
| $|\beta|^2$ | 0.533188 | 0.67207 | 0.77827 | | |

Table S3-c. The fitted parameters of the 2322-nm DPAVBi microbelt.

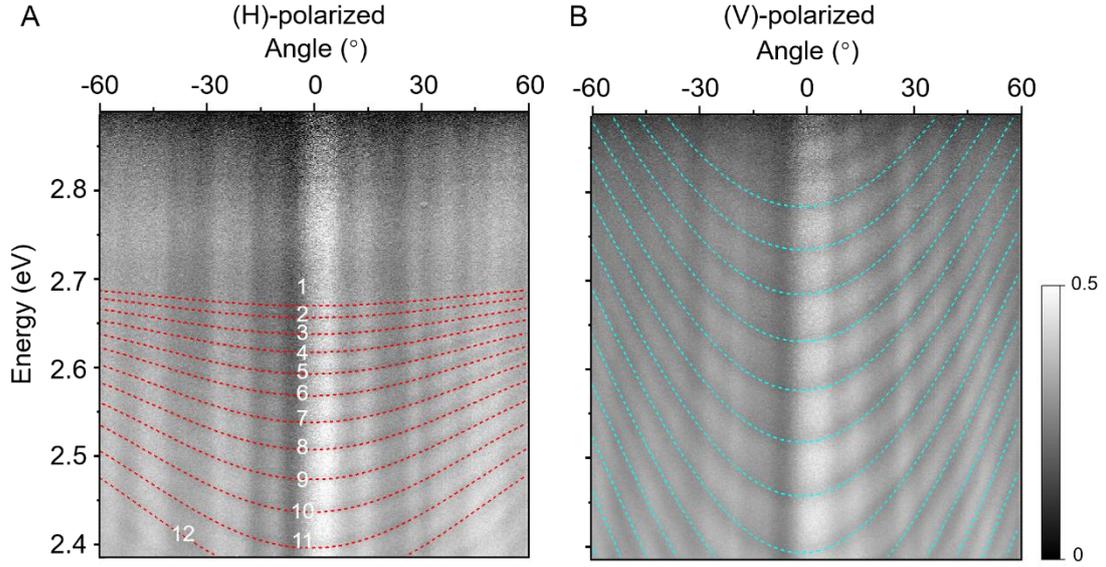

Figure S3-d. The angle-resolved reflectivity of the 4232-nm DPAVBi microbelt at (A) V-polarization coupled LP and (B) H-polarization uncoupled cavity mode. The Rabi splitting energy is fitted to be 538 meV.

| coupling mode | LP$_1$ | LP$_2$ | LP$_3$ | LP$_4$ | LP$_5$ |
|---|---|---|---|---|---|
| $E_{\theta=0°}$ (eV) | 2.67001 | 2.6551 | 2.63444 | 2.61211 | 2.58602 |
| detuning (meV) | 770 | 590 | 455 | 345 | 245 |
| $|\alpha|^2$ | 0.968169 | 0.954661 | 0.930005 | 0.893774 | 0.836031 |
| $|\beta|^2$ | 0.031831 | 0.045339 | 0.069995 | 0.106226 | 0.163969 |
| coupling mode | LP$_6$ | LP$_7$ | LP$_8$ | LP$_9$ | LP$_{10}$ |
| $E_{\theta=0°}$ (eV) | 2.56019 | 2.52919 | 2.49839 | 2.4655 | 2.43068 |
| detuning (meV) | 165 | 85 | 17 | -47 | -108 |
| $|\alpha|^2$ | 0.760778 | 0.650215 | 0.5315 | 0.414328 | 0.314367 |
| $|\beta|^2$ | 0.239222 | 0.349785 | 0.4685 | 0.585672 | 0.685633 |
| coupling mode | LP$_{11}$ | LP$_{12}$ | | | |
| $E_{\theta=0°}$ (eV) | 0.235895 | 0.182481 | | | |
| detuning (meV) | -168 | -222 | | | |

| | | |
|---|---|---|
| $\|\alpha\|^2$ | 0.235895 | 0.182481 |
| $\|\beta\|^2$ | 0.764105 | 0.817519 |

Table S3-d. The fitted parameters of the 4232-nm DPAVBi microbelt.

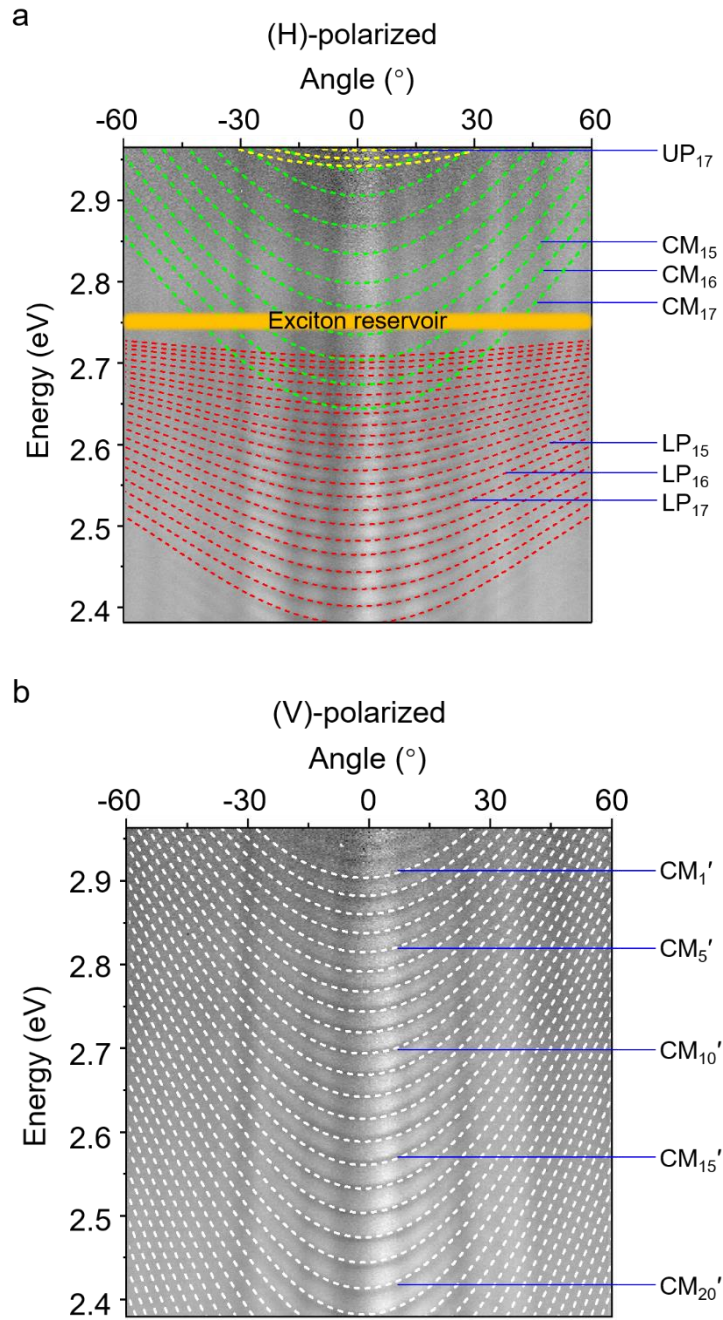

**Figure S3-e.** Measured and simulated angle-resolved reflectivity of the strong coupling of excitons at 2.754 eV and different cavity modes with a thickness around

8.6 μm. (a) ARR spectrum detected under polarization parallel to the length direction of the belt (defined as H-polarized), showing only the dispersions of lower polariton branches ($LP_n$) that are consistent with the coupled harmonic oscillator (CHO) model fitting

**Table S3-e.** CHO model fitting results for $LP_1$ to $LP_{20}$. The $E_{\theta=0°}$ represents the energy of $LP_n$ at $\theta = 0°$. The magnitudes $|\alpha|^2$ and $|\beta|^2$ correspond to the photonic and the excitonic fraction, respectively. The Rabi splitting energy is fitted to 512 meV.

| coupling mode | $LP_1$ | $LP_2$ | $LP_3$ | $LP_4$ | $LP_5$ | $LP_6$ | $LP_7$ | $LP_8$ |
|---|---|---|---|---|---|---|---|---|
| $E_{\theta=0°}$ (eV) | 2.715 | 2.705 | 2.693 | 2.681 | 2.672 | 2.658 | 2.647 | 2.637 |
| detuning (meV) | 525 | 460 | 395 | 340 | 300 | 250 | 210 | 180 |
| $|\alpha|^2$ | 0.948 | 0.935 | 0.918 | 0.897 | 0.878 | 0.847 | 0.814 | 0.785 |
| $|\beta|^2$ | 0.052 | 0.065 | 0.082 | 0.103 | 0.122 | 0.153 | 0.186 | 0.215 |
| coupling mode | $LP_9$ | $LP_{10}$ | $LP_{11}$ | $LP_{12}$ | $LP_{13}$ | $LP_{14}$ | $LP_{15}$ | $LP_{16}$ |
| $E_{\theta=0°}$ (eV) | 2.623 | 2.609 | 2.595 | 2.581 | 2.567 | 2.552 | 2.535 | 2.519 |
| detuning (eV) | 140 | 105 | 70 | 40 | 10 | -19 | -50 | -75 |
| $|\alpha|^2$ | 0.737 | 0.687 | 0.631 | 0.582 | 0.519 | 0.471 | 0.406 | 0.362 |
| $|\beta|^2$ | 0.263 | 0.313 | 0.369 | 0.418 | 0.481 | 0.529 | 0.594 | 0.638 |
| coupling mode | $LP_{17}$ | $LP_{18}$ | $LP_{19}$ | $LP_{20}$ | | | | |
| $E_{\theta=0°}$ (eV) | 2.506 | 2.488 | 2.471 | 2.454 | | | | |
| detuning (eV) | -105 | -132 | -158 | -184 | | | | |

| $|\alpha|^2$ | 0.313 | 0.274 | 0.241 | 0.212 |
| $|\beta|^2$ | 0.687 | 0.726 | 0.759 | 0.788 |

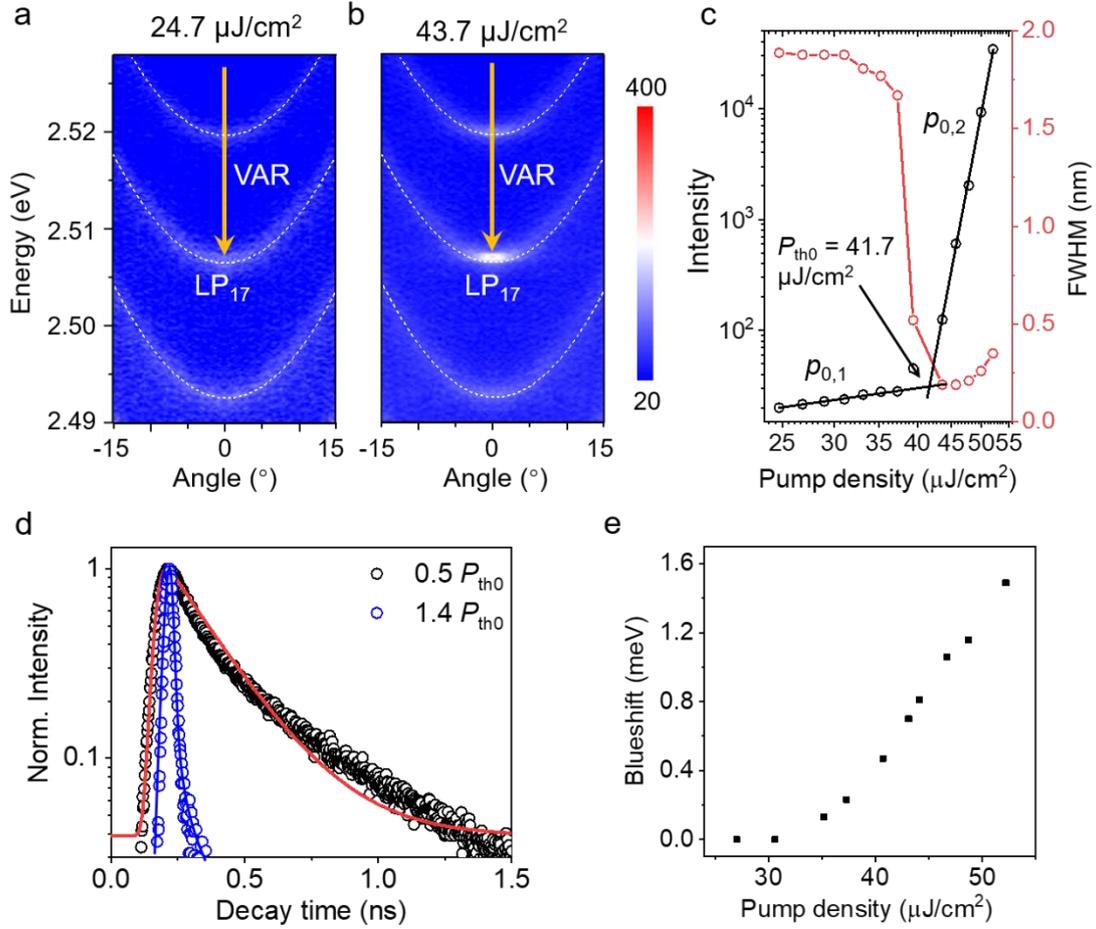

**Figure S4.** (a) (b) ARPL spectra at different pump densities. (c) The integrate PL intensity (black squares) and the line width (red squares) of the $LP_{17}$ as a function of pump density. (d) The PL decay at 0.5 $P_{th0}$ and 1.4 $P_{th0}$. (e) Energy blue shift as a function of pump power.

As shown in Fig. S4a, the APRL is consistent with the simulated polariton dispersions (white dotted lines) at a pump density below the threshold of 24.7 µJ cm$^{-2}$, proving that the ARPL indeed originates from polariton emissions. As the pumping density increases to 43.7 µJ cm$^{-2}$, the APRL (Fig. S4b) shows that polaritons condense at the bottom of $LP_{17}$ within the angle of around θ = ±2°, with the reduction of a full wide at half-maximum (FWHM) to 0.16 nm at 2.507 eV. The emission transition from sublinear to superlinear regime at an incident excitation density of ~41.7 μJ·cm$^{-2}$ ($P_{th0}$) is observed (Fig. S4c). The intensity dependence is separately fitted to a power law $x^p$ with $p_{0,1}$ = 0.69 ± 0.04 and $p_{0,2}$ = 31.7 ± 1.2, respectively. We also record the time-resolved PL from the LP$_{17}$ with a streak camera in Fig. S4d. At a very low pump density of 0.5 $P_{th}$, the polariton PL follows a single-exponential decay with a lifetime of τ = 0.197 ± 0.003 ns. Above the threshold, The lifetime (τ) of polariton PL decay collapses to <15 ps. These results confirm the macroscopic occupation at the ground state of LP$_{17}$, one of the main features of polariton condensation. According to the current research results, the vibration-assisted relaxation (VAR) of the exciton reservoir in organic semiconductors is the main channel for the buildup of the polariton population (orange arrows in Fig. S4a and b). An effective single-step relaxation is observed from the ARPLs, from the exciton reservoir selectively populates the LP$_{17}$ ground state, which coincide to the strongest vibronic progression with 0-1 band at 2.51 eV.

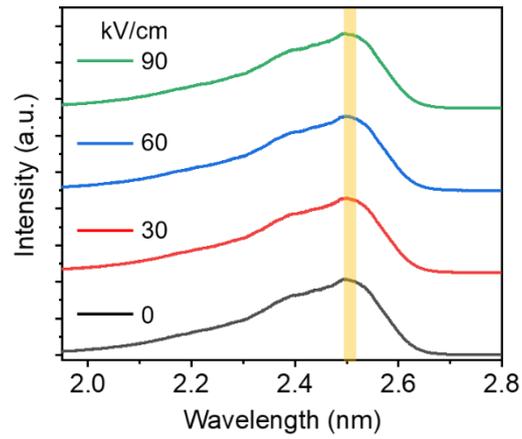

**Figure S5.** The exciton emission directly from the side of the sample at a fixed pump density (lower than the threshold $P_{th0}$) and the electric fields. With the increase of electric field, the exciton energy of the 0-1 exciton (yellow band) remains at 2.51 eV without any change.

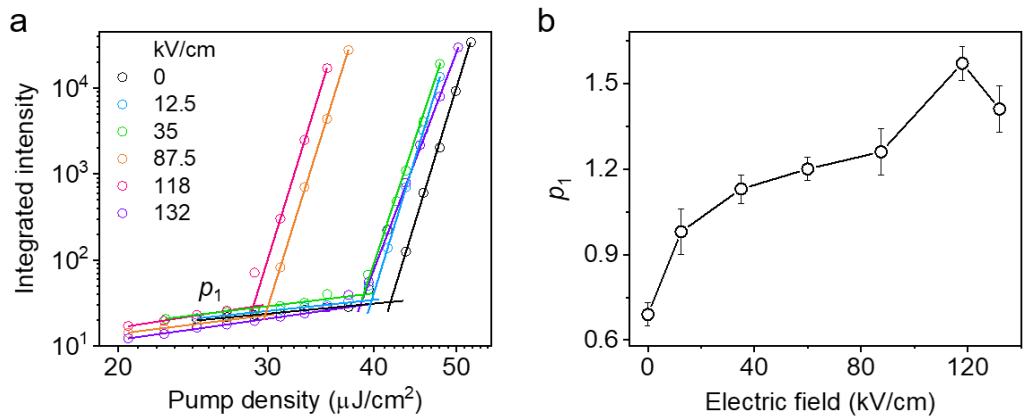

**Figure S6.** (a) The integrated PL intensity as a function of the pump density under different electric field. (b) The intensity dependence is separately fitted to a power law $x^p$, with $p_1$ below threshold under different electric field.

The parameter $p_1$ exhibits significant increase with the improvement of electric field, such as, $p_1$ increase from $0.69 \pm 0.04$ without electric field to $1.57 \pm 0.06$ under electric field of 118 kV/cm. When the electric field reaches 132 kV/cm, $p_1$ shows the significant decrease.

**Table S2.** The values of the parameters in Eqs. (1-2) in the main text used for the calculation of the curves in Fig. 3e.

| $n_{r0}$ | $\Gamma_{x0} = \Gamma_x$ | $\Gamma_p$ | $W_1$ | $W_2$ | $W_3$ | $E_{sat}$ |
|---|---|---|---|---|---|---|
| $10^4$ | $1 \times 10^{12}\ s^{-1}$ | $5 \times 10^{12}\ s^{-1}$ | $1.8 \times 10^4\ s^{-1}$ | $2.04 \times 10^3\ s^{-1}$ | $0.87 \times 10^3\ s^{-1}$ | 35 kV/cm |

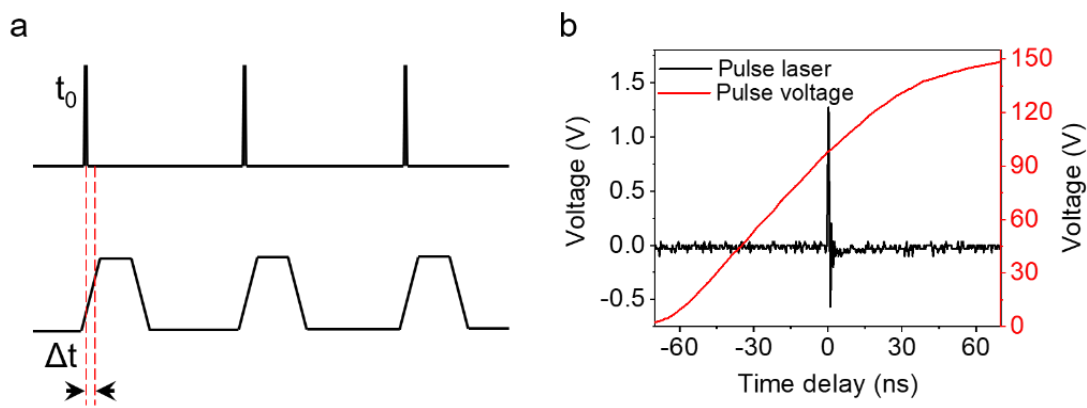

**Figure S7.** (a) Schematic diagram of laser pulse and electric pulse. (b) The laser pulse (dark line) and the rising edge of electrical pulse output by amplifier (red line) recoded by oscilloscope. The $t_0$ is defined as the moment when the laser pulse coincides with the rising electric pulse of 100V. The delay time (Δt) is controlled by delayed pulse generator. The rising speed of electric pulse is around 2.5 kV cm$^{-1}$ ns$^{-1}$.

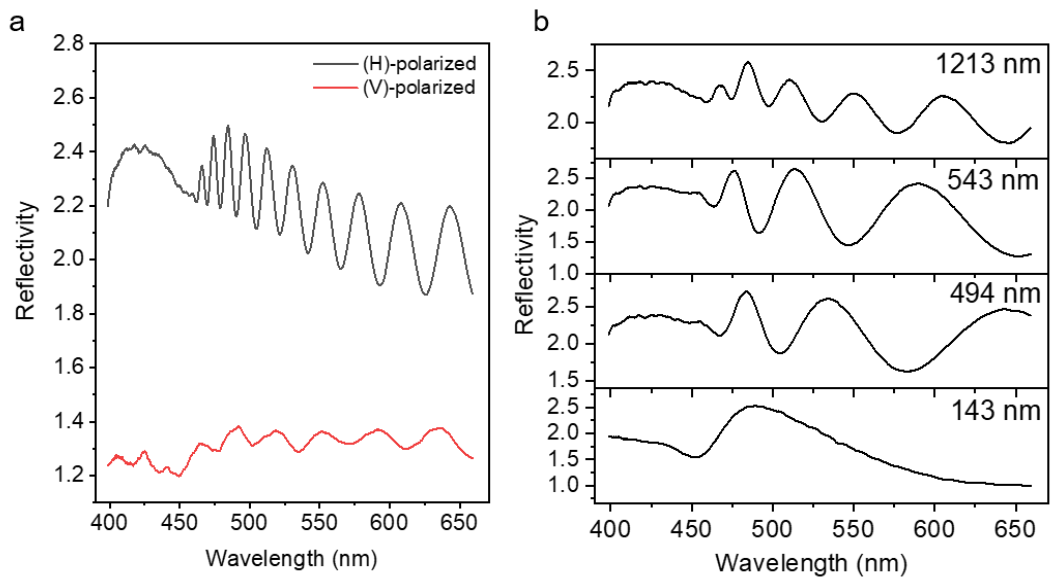

**Figure S8.** (a) Reflective spectra of the 1909-nm-microbelt cavity at the polarization of incident light parallel (black line) and vertical (red line) to the long direction of microbelts. (b) Reflective spectra of the organic cavity with the thicknesses of 1213 nm, 543 nm, 494 nm and 143 nm for parallel-polarization light.

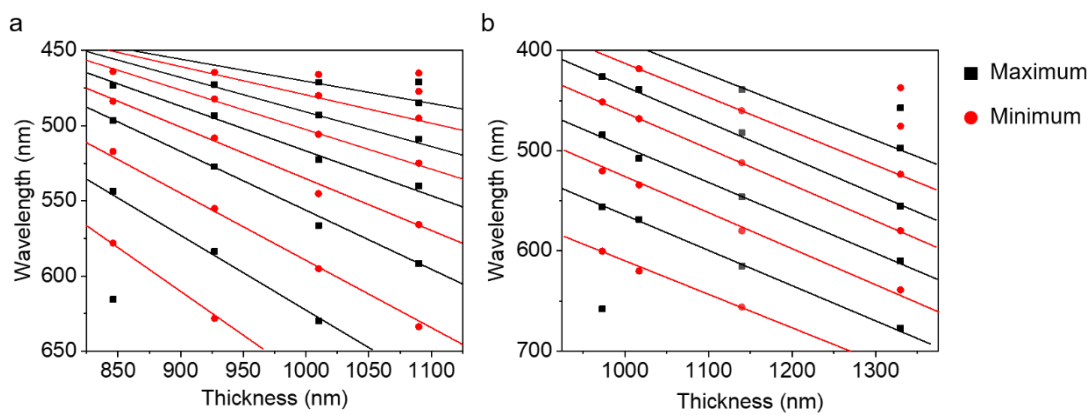

**Figure S9.** Wavelength of the interference maximum (black points) and minimum (red points) observed in reflection spectra measured by parallel- (a) and vertical-polarization light (b).

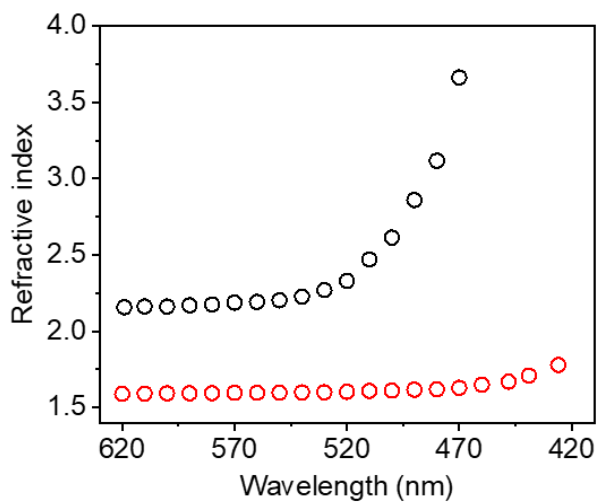

**Figure S10.** The calculated refractive indices in the direction parallel (black points) and vertical to long direction (red points) of DPAVBi microbelts.

We measured the refractive index using the method in the reference[1]. Firstly, the DPAVBi microbelts are placed on the quartz substrate, and the Fabry-Pérot interference is formed on the top and bottom smooth surfaces. The reflection spectra of the samples are measured with polarized light on parallel to and perpendicular to the long direction of the crystal. Figure S9A shows the reflection spectrum of the sample with the thickness of 1909 nm, where the black line and red line represent the reflection spectra parallel to and perpendicular to the long direction of the crystal, respectively. When the polarization of incident light parallels to the long direction of the crystal, the interference peaks become denser in the high energy band (black line). In contrast, the interference peak spacing in the vertical direction does not change significantly (red line). Figure S9B shows the interference spectra of the microbelts with the thickness of 1213 nm, 543 nm, 494 nm and 143 nm, respectively. The interference conditions are given by $2n(\lambda)d = m\lambda$, where $n(\lambda)$ is the refractive index at wavelength $\lambda$, $m$ is the order of interference, and $d$ is the crystal thickness. In the reflection spectra, the interference minimum occurs when $m$ is an integer and the maximum occurs when $m$ is a half integer.

The wavelengths of the interference maximum and minimum were extracted from the reflection spectra (Figure S9) measured when the polarization of incident light parallel to and vertical to the long direction of microbelts and are plotted as a function of crystal thickness (Figure S10A). The black points correspond to the maximum and the red points correspond to the minimum. The black (red) lines are fitted by the black (red) points. Since two adjacent black (or red) lines have difference 1 in the interference order, the order m can be determined by $m = d_1/(d_2-d_1)$, where $d_1$ and $d_2$ ($d_2 > d_1$) are the thicknesses corresponding to the order $m$ and $m + 1$, respectively, at a fixed wavelength $\lambda$. The calculated $n(\lambda)$ is shown in Figure S11. The $n(\lambda)$ in the direction parallel to the long direction of the microbelt rises from 2.157 at 620 nm to 3.664 at 470 nm, and rises from 1.592 at 620nm to 1.782 at 426 nm in vertical direction (Figure S10B). The rapid increase of $n(\lambda)$ near the exciton resonance strongly supports the conclusion of the strong exciton-photon coupling.